\documentclass[a4paper]{article}
\usepackage{url}
\usepackage{INTERSPEECH2020}
\usepackage{cite}
\usepackage{tablefootnote}
\usepackage{color}
\usepackage{graphicx} 
\usepackage{epstopdf, multirow}

\usepackage{pifont}
\usepackage{footnote}
\makesavenoteenv{tabular}

\usepackage{soul}
\title{Unsupervised Subword Modeling Using Autoregressive Pretraining and Cross-Lingual Phone-Aware Modeling}
\name{Siyuan Feng, Odette Scharenborg}
\address{
  Multimedia Computing Group, 
  Delft University of Technology, Delft,  the Netherlands}
\email{\{s.feng, o.e.scharenborg\}@tudelft.nl}

\begin{document}

\maketitle
\begin{abstract}
This study addresses unsupervised subword modeling, i.e., learning feature representations that can distinguish subword units of a language. 
The proposed approach adopts a two-stage bottleneck feature (BNF) learning framework, consisting of 
autoregressive predictive coding (APC) as a front-end  and a DNN-BNF model as a back-end. 
APC pretrained features are set as input features to a DNN-BNF model. A  language-mismatched ASR system is used to provide cross-lingual phone labels for DNN-BNF model training. Finally, BNFs are extracted as the subword-discriminative feature representation. A second aim of this work is to investigate the robustness of our approach's effectiveness to different amounts of training data. The results on Libri-light and the ZeroSpeech 2017 databases show that APC is effective in front-end feature pretraining.
Our whole system outperforms the state of the art on both databases.
Cross-lingual phone labels for  English data by a Dutch ASR outperform those by a Mandarin ASR, possibly linked to the larger similarity of Dutch compared to Mandarin with English.
Our system is less sensitive to training data amount  when the training data is over 50 hours.  
APC pretraining leads to a reduction of needed training material from over 5,000 hours to around 200 hours with little performance degradation.

\end{abstract}

\section{Introduction}

Training a DNN acoustic model (AM) for a high-performance automatic speech recognition (ASR) system requires a huge amount of speech data paired with  transcriptions. 
Many languages in the world have very limited or even no transcribed data \cite{dunbar2017zero}. 
Conventional supervised acoustic modeling techniques are thus problematic or even not applicable to these languages.

Unsupervised acoustic modeling (UAM) refers to the task of modeling basic acoustic units of a language with only untranscribed speech \cite{chen2015parallel,heck2017feature,Kamper2017segmental,Tjandra2019,Feng2019combining,Ondel2019Bayesian}. 
An important task in UAM is to learn frame-level feature representations that can distinguish subword units of the language for which no transcriptions are available, i.e., the \textit{target} language, and is 
robust to non-linguistic factors, such as speaker change \cite{versteegh2015zero,dunbar2017zero}. This problem is referred to as \textit{unsupervised subword modeling}, and is the focus of this study. It is essentially a feature representation learning problem.


There are many interesting attempts  to unsupervised subword modeling \cite{chen2015parallel,heck2017feature,chorowski2019unsupervised,shibata2017composite,feng2019_TASLP,riviere2020unsupervised,Feng2019combining}. One research strand is to use purely unsupervised learning techniques 
\cite{chen2015parallel,heck2017feature,chorowski2019unsupervised}. For instance, Chen et al. \cite{chen2015parallel} proposed a Dirichlet process Gaussian mixture model (DPGMM) posteriorgram approach, which performed the best in ZeroSpeech 2015 \cite{versteegh2015zero}.  Heck et al. extended this approach by applying unsupervised speaker adaptation, which performed the best in ZeroSpeech 2017 \cite{heck2017feature}. 
In a recent study \cite{Feng2019improving}, a two-stage bottleneck feature (BNF) learning framework was proposed.
The first stage, i.e., the front-end, used the factorized hierarchical variational autoencoder (FHVAE) \cite{hsu2017nips} to learn speaker-invariant features. The second stage, the back-end, consisted of a DNN-BNF model \cite{chen2017multilingual}, which used the FHVAE pretrained features as input features and generated BNFs as the desired subword-discriminative acoustic feature representations. In the case of unsupervised acoustic modeling, no frame labels are available for DNN-BNF model training.
In \cite{Feng2019improving},
 DPGMM was adopted  
 as a building block of the back-end
 to generate pseudo-phone labels for the speech frames.
In another recent study  \cite{chorowski2019unsupervised}, the vector quantized VAE (VQ-VAE) \cite{oord2017neural} was applied to directly learn the desired feature representation without a back-end model such as the DNN-BNF, and is comparable to state-of-the-art performance \cite{heck2017feature}. 

In another research strand, frame-level feature representations that can distinguish subword units in the target language are created using a cross-lingual knowledge transfer approach \cite{feng2019_TASLP,shibata2017composite}. Here, out-of-domain (OOD) mismatched language resources are used to train DNN AMs which are further used to extract phone posteriorgrams or BNFs of the target speech.
The two research strands mentioned above can also be combined.
For instance, \cite{feng2019_TASLP} proposed to 
apply the DNN-BNF model, and  utilized  unsupervised   DPGMM  and OOD ASR systems  to  generate two types of frame labels for multi-task DNN-BNF learning. The two label types correspond to  the two research strands respectively. 
The results  showed the complementarity of the two label types in unsupervised subword modeling.

The present study  adopts a two-stage BNF learning framework similar to  \cite{Feng2019improving}, and aims at
combining unsupervised   learning techniques, specifically autoregressive predictive coding (APC) as a front-end, with  cross-lingual   knowledge transfer in the back-end. 
Recently, APC   has   been shown   \cite{Chung2019} to learn speech feature representations that are beneficial to 
various downstream   tasks, and outperform other effective
unsupervised methods such as contrastive predictive coding (CPC) \cite{oord2018cpc} 
in ASR, speech translation and speaker verification   \cite{Chung2019generative}.
 APC preserves   phonetic (subword) and speaker information from the original speech signal, while the two information types are more  separable. This makes APC a possibly interesting  method for unsupervised subword modeling. 
In this paper, we investigate the effectiveness of APC in this task for the first time.

In  the second stage, a DNN-BNF back-end is trained, using the APC pretrained features as input features. 
Frame labels required for  DNN-BNF model training are obtained using an OOD  ASR system as was done in \cite{feng2019_TASLP}.  
By doing so, cross-lingual phonetic knowledge is exploited.
Two OOD ASR systems trained on  different OOD languages are employed for comparison, in order to study the effect of target and OOD language similarity on the performance of the proposed approach.


For low-resource languages for which transcribed data are absent, 
even unlabeled speech can be costly to collect. The robustness of unsupervised subword modeling methods
against limited amounts of training material is therefore an important topic, however has received little attention in the literature so far. The second aim of this work is therefore to systematically investigate the robustness 
of the proposed approach's effectiveness to different amounts of training data. Specifically, we varied the amount of training data from $10$ hours to over $500$ hours.
 

\section{Proposed approach}

\begin{figure}
    \centering
    \includegraphics[width=0.95\linewidth]{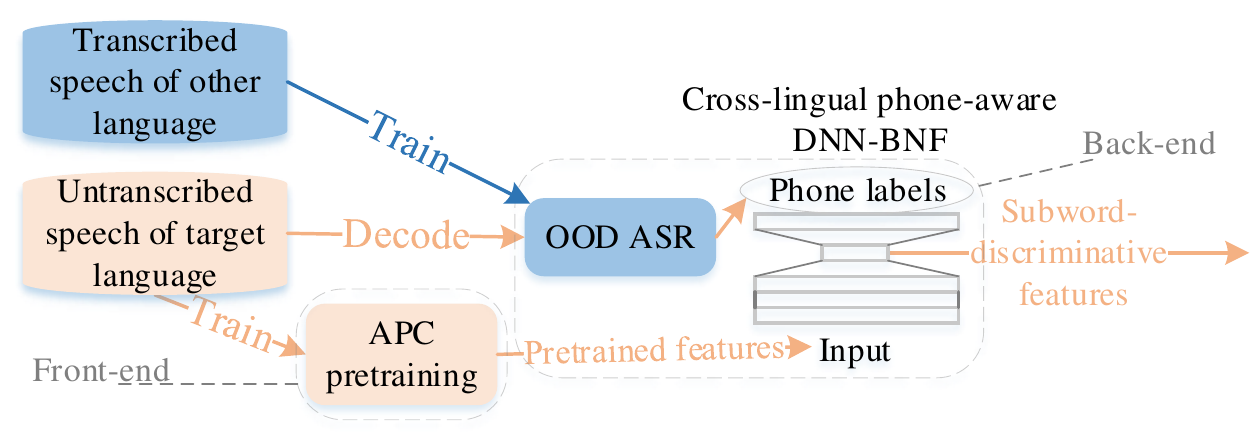}
    \caption{General framework of the proposed approach to unsupervised subword modeling.  }
    \label{fig:general_framework}
\end{figure}
The general framework of our proposed approach is illustrated in Figure \ref{fig:general_framework}. Given untranscribed speech data of a  target  language, an APC  model is pretrained in the front-end.
Next, an OOD ASR system trained on a language different from the target language  assigns a phone label to every frame of the target language's speech data through decoding. Pretrained features  created by the APC model and the cross-lingual phone labels created by the OOD ASR are then used to train a DNN-BNF model in the back-end, from which BNFs are extracted as the subword-discriminative representation in the final step.

Front-end APC pretraining will be compared with an FHVAE approach \cite{hsu2017nips} which was used in related previous work \cite{Feng2019improving}. The whole pipeline of our approach will be compared with a system consisting of only the back-end DNN-BNF model, and a CPC approach \cite{oord2018cpc} applied in the same task \cite{kahn2019librilight}.
Moreover, two different languages will be used to train two different OOD ASR systems for comparison.
\subsection{APC pretraining}
\label{subsec:apc}
In our concerned task, previously adopted feature learning methods 
usually target suppressing speaker variation, such as FHVAE \cite{Feng2019improving} and speaker adaptation \cite{heck2017feature}. In contrast, 
APC aims at learning
a representation that keeps information from speech, while phonetic information is made more separable from speaker information. The learned representation is considered less risky of losing phonetic information than 
  representations learned by methods in \cite{Feng2019improving,heck2017feature}.


Let us assume a set of unlabeled speech frames $\{\bm{x_1}, \bm{x_2}, \ldots, \bm{x_T}\}$ for training, where $T$ is the total number of frames. At each time step $t$, the encoder of APC model $\textrm{Enc} (\cdot)$ reads as input  a feature vector 
$\bm{x_t}$, 
and outputs  a feature vector  $\bm{\hat{x}_t}$ (same dimension as $\bm{x_t}$) based on all the previous inputs $\bm{x_{1:t}}=\{\bm{x_1},\ldots,\bm{x_t}\}$,
\begin{equation}
    \bm{\hat{x}_t} = \textrm{Enc} (\bm{x_{1:t}}).
    \label{eqt:enc}
\end{equation}
The goal of APC is to let $\bm{\hat{x}_t}$ be as close as possible to $\bm{x_{t+n}}$, where $n$ is a pre-defined constant positive integer, denoted as \textit{prediction step}. 
The loss function during APC   training is defined as: $\textrm{Loss} = \sum_{t=1}^{T-n} \left| \bm{\hat{x}_t} - \bm{x_{t+n}} \right|$.
Intuitively, increasing $n$ encourages the   encoder to capture contextual dependencies in speech, 
while a small $n$ focuses more on local smoothness.
 
Here, the encoder of APC $\mathrm{Enc} (\cdot)$ is realized by a long short-term memory (LSTM) \cite{hochreiter1997long} RNN. Let $L$ denote the number of LSTM layers,  Equation (\ref{eqt:enc}) is formulated as,
\begin{align}
    \bm{h_0} &= \bm{x_{1:t}},    \\
    \bm{h_l} &= \textrm{LSTM}^l (\bm{h_{l-1}}), l\in \{1,2,\ldots, L\}, \\
    \bm{\hat{x}_t} &= \bm{W} \bm{h_L},
\end{align}
where $\bm{W}$ is a trainable projection matrix. The equations that form  $\textrm{LSTM} (\cdot)$ can be found in \cite{sak2014long}. 

After APC training, the output of the top hidden layer $\bm{h_L}$ is extracted as the  learned acoustic representation, and is henceforth referred to as the  \textit{APC feature}. 
Although in principle,  $\bm{h_l}$ of any layer $l$ could be used as the learned representation, 
we follow \cite{Chung2019} in using the output of the top layer as they showed that this gave the best results in phone classification tasks.

\subsection{Cross-lingual phone-aware DNN-BNF}
As shown in Figure \ref{fig:general_framework}, the DNN-BNF back-end
is a DNN with a bottleneck   layer in the middle \cite{grezl2009investigation}. 
To obtain cross-lingual phone labels, the OOD ASR is used to decode  target speech utterances into  lattices,  and find the best path for every utterance. Afterwards, each speech frame is assigned with a triphone HMM state modeled by the OOD ASR. These   state labels provide phonetic representation for the target speech from a cross-lingual perspective. 


After obtaining triphone HMM state labels  as cross-lingual phone labels, the DNN-BNF  is trained using the pretrained APC features and the cross-lingual phone labels in a supervised manner \cite{grezl2007probabilistic}, and used to  extract BNFs   as the desired subword-discriminative feature representation. 

\section{Experimental setup}

\subsection{Databases and evaluation metric}
 English is chosen as the target language while Dutch and Mandarin are chosen as the two OOD languages. 
Training data for APC pretraining and DNN-BNF model training are taken from  Libri-light \cite{kahn2019librilight}, 
a newly published English database
to support unsupervised subword modeling. 
The \textit{unlab-600} and  \textit{unlab-6K} sets from Libri-light are adopted. Unlab-600 is used in both APC pretraining and DNN-BNF model training, while unlab-6K is used only in DNN-BNF model training.
Unlab-600   consists of 
$526$ hours of speech excluding silence.
Additionally, we randomly select four subsets of utterances from unlab-600 to investigate the robustness of our approach to different amounts of training material. These subsets consist of $900$ (i.e., $13$ hours), $3.6$K ($52$ hours), $7.2$K ($104$ hours), and $14.4$K ($209$ hours)   utterances. 
Unlab-6K  set consists of $5,273$ hours of
speech excluding silence.  Details of the training sets are listed in Table \ref{tab:libri_light_data}.
\begin{table}[!t]
\renewcommand\arraystretch{0.7}
\centering
\caption{Libri-light training data and its subsets.}
\resizebox{ 0.9 \linewidth}{!}{%
\begin{tabular}{l|c|c|cccc}      
\toprule
&unlab-6K& unlab-600 & \multicolumn{4}{c}{subsets of unlab-600} \\
\midrule
\#utterances &$362,817$ &  $36,229$ & $14,400$ &$7,200$ &$3,600$ &$900$ \\
\#speakers & $1,742$ & $489$ & $438$ & $393$ & $351$ & $244$ \\
Hours  & $5,273$ &$526$& $209$ & $104$ & $52$ & $13$ \\
\bottomrule
\end{tabular}%
}
\label{tab:libri_light_data}
\end{table}

The Dutch and Mandarin corpora used for training the two OOD ASR systems are the CGN \cite{oostdijk2000spoken} and Aidatatang\_200zh \cite{aidatatang}, respectively. 
The CGN   training and test data partition follows \cite{laurensw75cgn_kaldi}. Its training data contains $483$ hours of speech, covering speaking styles including conversational and read speech and broadcast news.
Aidatatang\_200zh is a read speech   corpus.
Its training data contains $140$ hours of speech.   

Evaluation data are taken from  Libri-light and ZeroSpeech 2017 \cite{dunbar2017zero}. 
 Libri-light evaluation sets consist of \textit{dev-clean, dev-other, test-clean} and \textit{test-other}.
 , with $*$-clean  having higher recording quality and accents closer to US English than   $*$-other
 \cite{panayotov2015librispeech}. They are used to evaluate the effectiveness of  both front-end pretrained features and  BNFs learned by the back-end.

Evaluation on ZeroSpeech 2017 aims  to better compare our approach with previous research in this area.
English evaluation data from ZeroSpeech 2017 are used to evaluate the effectiveness of BNFs learned by the the back-end. These data are organized into subsets of differing   lengths (1s, 10s \& 120s)  \cite{dunbar2017zero}.


The created BNFs, as well as APC pretrained features, are evaluated in terms of the ABX subword discriminability \cite{dunbar2017zero}.
In the ABX task, $A$, $B$ and $X$ are three speech segments, and $x$ and $y$ are two different phonemes. $A \in x$, $B \in y$, $X \in x$ or $y$. Following \cite{dunbar2017zero} (see also for more details), an error occurs if given a pre-defined distance measure $d$, $d(A,X) > d(B,X)$, given $X \in x$, 
or $d(A,X) < d(B,X)$, given $X \in y$.
Dynamic time warping is chosen as the distance measure. Segments $A$ and $B$   belong to the same speaker. ABX error rates for \textit{within-speaker} and \textit{across-speaker} are evaluated separately, depending
on whether X and A belong to the same speaker.

\subsection{Front-end}
\label{subsec:exp_setup_apc_fhvae}
The APC model is implemented as a  multi-layer  LSTM network. Residual connections are made between two consecutive layers.  
Each LSTM layer has $100$ dimensions.
Unless specified explicitly, the number of LSTM layers is $3$.
For each training data amount setting,
the
prediction step $n$ (in Section \ref{subsec:apc}) is
picked from $\{1,2,3,4,5\}$ which gives the best ABX performance.
Our preliminary experiments showed that increasing $n$ to larger than $5$ would lead to rapid degradation in ABX error rate. 
The input features to APC are $13$-dimension MFCCs with cepstral mean normalization (CMN) at speaker level. 
The model is trained with the open-source tool by \cite{Chung2019} for $100$ epochs with the Adam optimizer \cite{kingma2014adam}, an initial learning rate of $10^{-4}$, and a batch size of $32$. 
After training, the top LSTM layer's output is extracted as the APC feature representation.

The performance of front-end  APC pretraining is compared against FHVAE \cite{hsu2017nips}, which was used in related previous work \cite{Feng2019improving}. 
The latent representation $\bm{z_1}$ of FHVAE is known to be preserving linguistic content while suppressing speaker variation \cite{hsu2017nips}, and is compared with the APC feature representation. The model architecture of FHVAE and its training procedure follow those in \cite{Feng2019improving}.
The FHVAE models are trained 
using an open-source tool \cite{hsu2017nips}, and take the same input features and training data (i.e., Libri-light) as the APC models. After training, the  FHVAE encoder's output $\bm{z_1}$ is extracted.
\subsection{Back-end}
\subsubsection{OOD ASR systems}
\label{subsubsec:oodasr}
We trained two OOD ASR systems, i.e., a Dutch ASR and a Mandarin ASR.
The OOD ASR systems use a chain-time delay NN (TDNN) AM \cite{povey2016purely} trained using Kaldi \cite{povey2011kaldi}, containing $7$   layers.  The TDNN   is trained based on the lattice-free maximum mutual information (LF-MMI) criterion \cite{povey2016purely}.
For Dutch, the input features consist of $40$-dimension high-resolution (HR) MFCCs. For Mandarin, the input  features consist of HR MFCCs  appended by 
pitch features \cite{ghahremani2014pitch}. 
Frame labels required for TDNN training are obtained by forced-alignment with a GMM-HMM AM trained beforehand. 
 For both systems, a tri-gram LM is trained using training data transcriptions.

The Dutch ASR obtained a word error rate (WER) of $8.98\%$ on the CGN broadcast test set. 
(This   WER could be improved upon by integrating an   RNN LM. However, as Dutch ASR performance is not the focus of this study, an RNN LM is not applied.)
The Mandarin ASR obtained a character error rate (CER) of $6.37\%$ on the Aidatatang\_200zh test set.
The two ASR systems are used to generate cross-lingual phone labels for Libri-light training speech frames.




\subsubsection{DNN-BNF setup}
Two DNN-BNF models are trained, one taking the Dutch cross-lingual phone labels as training labels and one taking the Mandarin  phone labels as training labels. 

The DNN-BNF consists of $7$ feed-forward layers (FFLs). Each layer has $450$ dimensions except a $40$-dimension bottleneck layer, which is located below the top FFL. 
The DNN-BNF uses a chain model \cite{povey2016purely} which is trained based on  the   LF-MMI criterion. 
The inputs to DNN-BNF are the APC  feature with its neighboring ($-3$ to $+3$) frames. 
After DNN-BNF training, $40$-dimension  BNFs are extracted as the learned subword-discriminative representation and evaluated with the ABX task.


For the purpose of   comparison, two more DNN-BNF models are  trained using the $40$-dimension HR MFCC with its neighboring ($-3$ to $+3$) frames as input features. One model takes the Dutch labels  and the other takes the Mandarin labels.
Other  training and model settings   are unchanged.
After  training, BNFs are extracted and also evaluated with the ABX task.

\section{Results and discussion}
\subsection{Effectiveness of APC features }
\label{subsec:results_apc}

\begin{figure}[!t]
    \centering
    \includegraphics[width=0.95\linewidth]{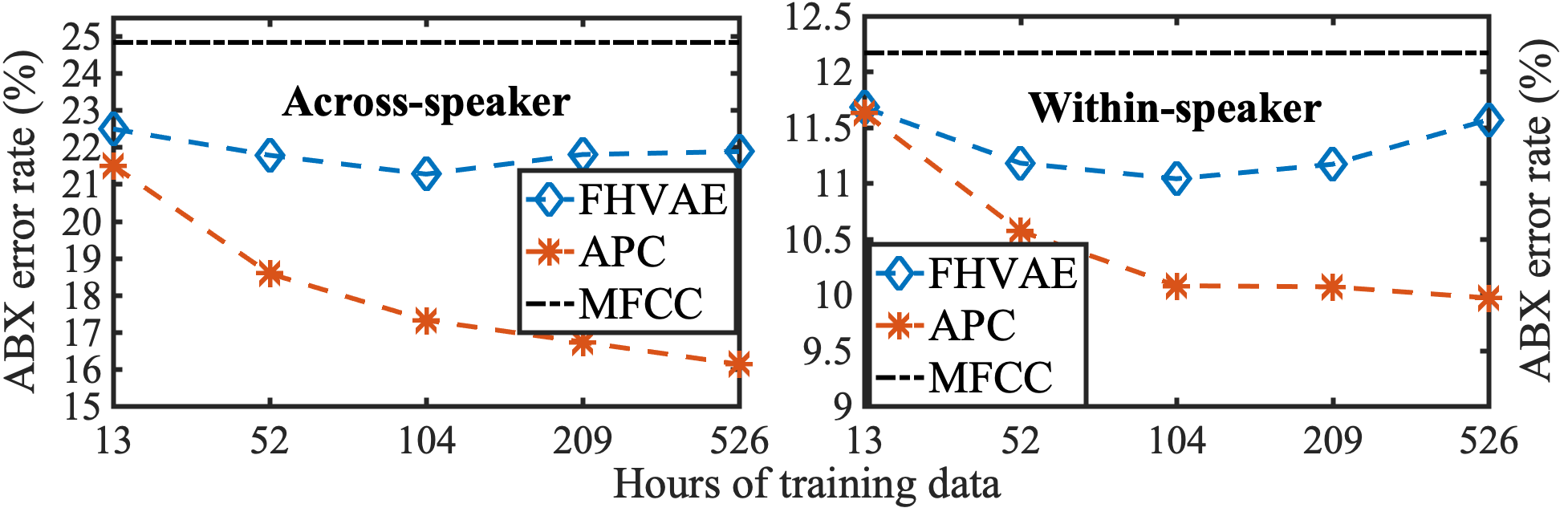}
    \caption{ABX error rates  of APC features, FHVAE features and official MFCC baseline on Libri-light (Avg. over $4$ sets). }
    \label{fig:apc_fhvae_mfcc}
\end{figure}

In this subsection, the APC  features and FHVAE features in the front-end ($\bm{z_1}$)  are directly evaluated using the ABX task, without being modeled by the  DNN-BNF back-end. 
ABX error rates ($\%$) of the APC     and   FHVAE features with respect to different hours of training data  are shown in Figure \ref{fig:apc_fhvae_mfcc}.
ABX results in this figure are averaged values over the $4$ evaluation sets in Libri-light. The official MFCC baseline   \cite{kahn2019librilight} is also shown in this figure. It can be observed that both the APC features and the FHVAE features  outperform the MFCC features. The APC features are consistently superior to the FHVAE features in both the across- and the within-speaker conditions irrespective of the amount of training data.
Figure \ref{fig:apc_fhvae_mfcc} (left) indicates that the APC features are more robust to speaker variation than the FHVAE features, even though the APC model is not explicitly suppressing speaker variation as FHVAE is.


\subsection{Effectiveness of BNF representation}
\label{subsec:results_bnf_librilight}
\begin{table}[!t]
\renewcommand\arraystretch{0.55}
\centering
\caption{ABX error rates of  BNFs, APC and CPC    on Libri-light.  Models are trained with unlab-600. APC   has $5$ layers, as this gives the best performance among different nos. of APC layers.}
\resizebox{0.96 \linewidth}{!}{%
\begin{tabular}{l|cc|cccc|c}      
\toprule
\midrule[0.2pt]
& \multicolumn{7}{c}{Across-speaker}\\
Feature&Input & Label & dev-clean & dev-other & test-clean & test-other & Avg.\\ 
\midrule
\multirow{4}{*}{BNF} &APC & Du & $\bm{6.18}$&	$\bm{11.02}$&$	\bm{6.03}$	&$\bm{10.94}$&$	\bm{8.54}$\\
&MFCC & Du &$6.67$&$	11.65$&	$6.64$&$	12.00$&$	9.24$\\
&APC & Ma & $7.00$&$11.80$&$6.84$&$11.81$&$9.36$\\
&MFCC & Ma & $7.92$&	$12.71$&	$7.74$&	$13.23$&	$10.40$\\
\midrule
APC& \multicolumn{2}{c|}{-} & $12.64$&$	19.00$&$12.19$&$18.75$&$15.65$\\
CPC \cite{kahn2019librilight} & \multicolumn{2}{c|}{-}&$9.58$&	$14.67$&	$9.00$&	$15.10$&	$12.09$\\

\midrule
&\multicolumn{7}{c}{Within-speaker} \\
\midrule
\multirow{4}{*}{BNF} &APC & Du  & $\bm{4.77}$&	$\bm{6.69}$&	$\bm{4.49}$&	$\bm{6.43}$&	$\bm{5.60}$\\
&MFCC & Du & $4.97	$&$6.94$&	$4.73$&	$6.86$&	$5.88$\\
&APC & Ma  & $5.25$&	$7.14$&	$5.21$&	$7.09$&	$6.17$\\
&MFCC & Ma & $6.06$&	$7.71$&	$5.62$&	$7.82$&	$6.80$\\
\midrule
APC&\multicolumn{2}{c|}{-}& $8.83$&$11.07$&$8.36$&$11.48$&$9.94$\\
CPC \cite{kahn2019librilight}&\multicolumn{2}{c|}{-} &$7.36$&	$9.39$&	$6.90$&	$9.59$&	$8.31$\\
\midrule[0.2pt]
\bottomrule
\end{tabular}%

}
\label{tab:dnn_bnf}
\end{table}
In this subsection, all models are trained with unlab-600 ($526$ hours).
ABX error rates (\%) of BNFs extracted by the back-end DNN-BNF model are listed in Table \ref{tab:dnn_bnf}. The second and third columns  denote input feature types and frame labels for training DNN-BNF models. `Du' and `Ma' stand for Dutch and Mandarin. 
Two front-end features, i.e. APC and CPC (in \cite{kahn2019librilight}),  are also listed as references.
   From this table, it is observed  that:

(1) DNN-BNF trained with APC features performs better than that trained with MFCC features in all the evaluation sets. This  demonstrates the effectiveness of front-end APC pretraining in our proposed two-stage system framework.

(2) The BNFs obtained from the back-end DNN-BNF model outperform the APC features from the front-end. In other words, the results show that
back-end DNN-BNF modeling with cross-lingual phone labels
outperforms front-end  pretrained features for unsupervised subword modeling, similar to what has been observed by \cite{shibata2017composite,feng2019_TASLP}. BNF also performs better than the   CPC feature    \cite{kahn2019librilight}. Note that CPC does not require OOD resources during training while BNF in this study does. 

(3) The performance achieved by adopting Dutch labels in DNN-BNF model training is slightly better than that by adopting Mandarin labels. 
This can possibly be explained by the similarity between the OOD language and target in-domain language, i.e., Dutch and English, respectively, which are both West Germanic languages, while Mandarin is not. 
Although one could 
possibly 
attribute the superiority of adopting Dutch labels over Mandarin labels to the larger amount of training data for Dutch ($483$ hours) than for Mandarin ($140$ hours), this is not a likely explanation because both models achieved fairly similar 
results 
on their respective in-domain test sets (in Section \ref{subsubsec:oodasr}).

Table \ref{tab:dnn_bnf} also shows CPC outperforms  APC. We plan to replace the front-end APC  with CPC and study its efficacy in combination with the back-end DNN-BNF model in the future. 








\subsection{Effect of amount of training data}
\begin{figure}[!t]
    \centering
    \includegraphics[width=0.91\linewidth]{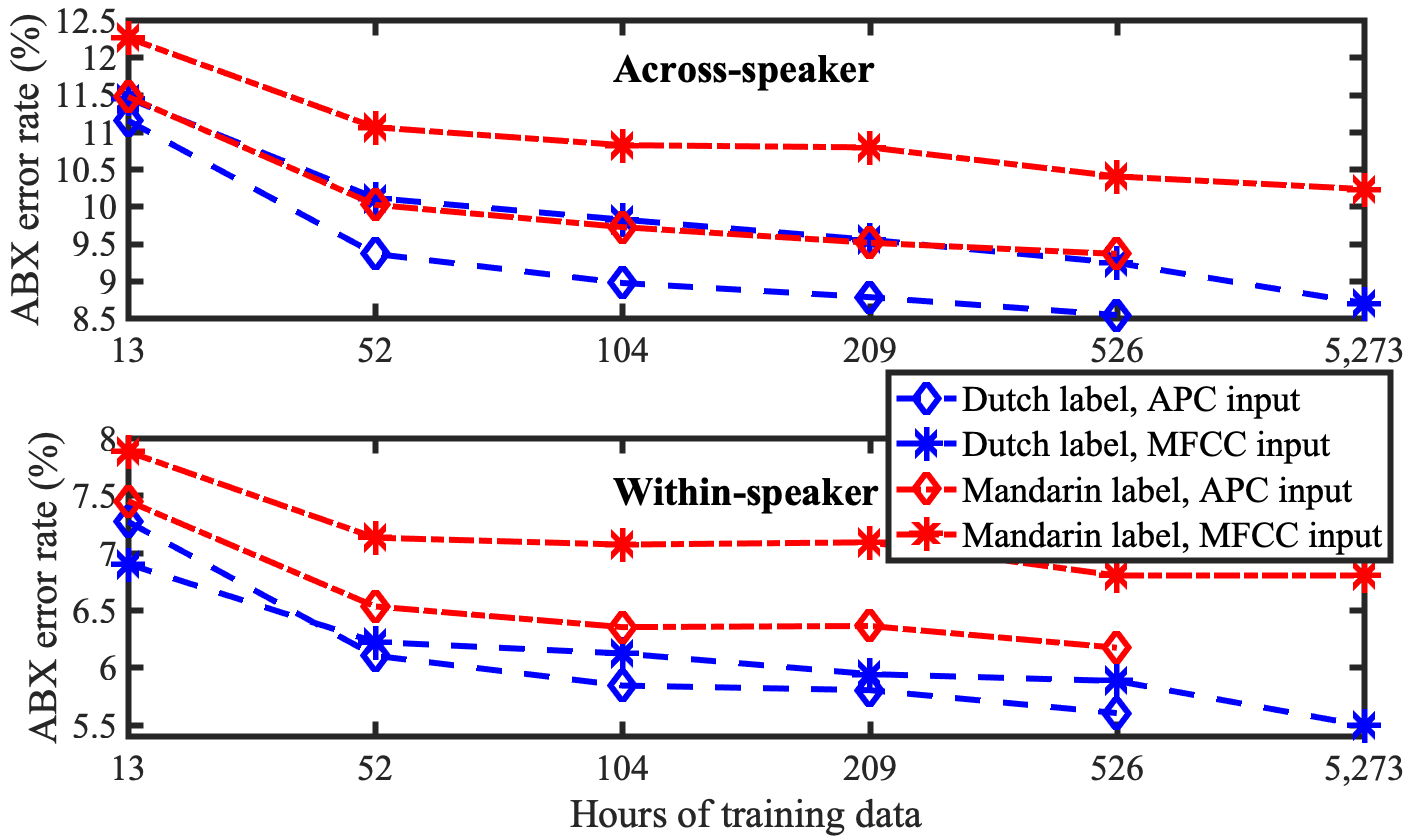}
    \caption{ABX error rates of BNFs w.r.t amount of training data on Libri-light (Avg. over $4$ sets).}
    \label{fig:dnn_bnf_data_amount}
\end{figure}

ABX error rates ($\%$) of BNFs extracted by  DNN-BNF models with respect to different amounts of training data in hours are illustrated in Figure \ref{fig:dnn_bnf_data_amount}. The results are averaged values over the $4$ evaluation sets in  Libri-light. Unlab-6K ($5,273$ hours)  is only adopted in training DNN-BNF models with MFCC input features (marked as ``$\ast$''). For models trained with APC features as input features (``$\diamond$''), the  data amount for APC pretraining and DNN-BNF model training   is the same for each run.  From Figure \ref{fig:dnn_bnf_data_amount}, it can clearly be seen that 
performance improves as more training data is available, with the largest improvement when the training data increases from $13$  hours  to $52$ hours, and less improvement for any additional training material.

Secondly, across the different data amounts, the DNN-BNF models trained with APC features as input features are almost consistently better than those with MFCC input features. 
Interestingly, 
with Dutch labels,
the model that uses APC features and is trained with $209$ hours of data achieves a similar across-speaker error rate ($8.78\%$) to the model trained with MFCCs with $5,273$ hours of data ($8.70\%$). This implies that APC pretraining ``saves'' around $5,000$ hours (i.e.  $96\%$) of training data, making APC pretraining highly appealing in low-resource speech modeling. The effect of pretraining on the needed amount of training data is even larger when Mandarin labels are used 
(saving over $99\%$ of the training data).




\subsection{ZeroSpeech 2017 results}



\begin{table}[!t]
\renewcommand\arraystretch{0.65}
\centering
\caption{ABX error rates of  BNFs  on ZeroSpeech 2017 English evaluation sets. Models are trained with  Libri-light.}
\resizebox{1 \linewidth}{!}{%
\begin{tabular}{cc|ccc|c|ccc|c}      
\toprule
\midrule[0.2pt]
& &\multicolumn{4}{c|}{Across-speaker}& \multicolumn{4}{c}{Within-speaker}\\
System &   Hours &  1s & 10s & 120s & Avg. &1s &10s &120s &Avg.\\
\midrule[0.2pt]

\multirow{ 3}{*}{Proposed-Du} & $526$  &  $7.65$&$6.69$&$6.66$&$\bm{7.00}$&$5.52$&$4.77$&$4.68$&$\bm{4.99}$\\
& $209$&  $8.11$&$6.99$&$6.90$&$7.33$&$5.83$&$5.06$&$4.97$&$5.29$\\
& $104$&  $8.14$&$7.07$&$7.03$&$7.41$&$5.89$&$4.99$&$5.00$&$5.29$ \\
\midrule
\multirow{ 3}{*}{Proposed-Ma} 
& $526$  &  $8.19$&$7.33$&$7.30$&$7.61$&$5.97$&$5.39$&$5.37$&$5.58$ \\
& $209$& $8.62$&$7.61$&$7.52$&$7.92$&$6.31$&$5.52$&$5.60$&$5.81$\\
& $104$&   $8.47$&$7.62$&$7.52$&$7.87$&$6.13$&$5.49$&$5.44$&$5.69$\\
\midrule[0.2pt]
Topline \cite{dunbar2017zero}&  &$8.6$&$6.9$&$6.7$&$7.40$ &$6.5$&$5.3$&$5.1 $&$5.63$\\
  \cite{shibata2017composite}& & $7.9$&$7.4$&$6.9$&$7.40$&$5.5$&$5.2$&$4.9$&$5.20$\\
\midrule[0.2pt]
\bottomrule
\end{tabular}%
}
\label{tab:zrsc2017}
\end{table}
We also evaluated the performance of  our approach 
on the ZeroSpeech 2017 English evaluation sets. The results are shown in Table \ref{tab:zrsc2017}, which also includes the official topline \cite{dunbar2017zero}  and the best-performing system (using OOD data) \cite{shibata2017composite}. 
Note that, unlike our approach, these two systems employed English labeled data. 
The total amount of labeled training data used in \cite{shibata2017composite} is $1,327$ hours (including $80$-hour English data).
In this table, ``Proposed-Du/-Ma'' denotes our proposed approach by adopting Dutch or Mandarin labels  respectively.
Interestingly, using Dutch labels, our system trained with $526$ hours of data outperforms the topline and \cite{shibata2017composite} systems, and is comparable to the two reference systems when trained with only $104$ hours of data. Table \ref{tab:zrsc2017} also shows the proposed approach by adopting Dutch labels performs better than that by adopting Mandarin labels,  which is consistent with observations in Section \ref{subsec:results_bnf_librilight}.



\section{Conclusions}
This study addresses unsupervised subword modeling, and proposes a two-stage system that consists of APC pretraining and cross-lingual phone-aware DNN-BNF modeling.  
Experimental results on Libri-light and ZeroSpeech 2017 databases demonstrate the effectiveness of APC in front-end feature pretraining. It surpasses a previously adopted FHVAE approach. Our whole system outperforms the state of the art on both databases. 
Cross-lingual phone labeling for English data by a Dutch ASR is slightly better than by a Mandarin ASR. This is possibly linked to the larger similarity of Dutch than Mandarin with  English.  
The proposed approach benefits from increasing training data amount, and is less sensitive to data amount when the training data is over $50$ hours. 
When using APC pretraining, $4\%$ of the training material could result in a similar   performance to   using the full training set without  APC pretraining.


\bibliographystyle{IEEEtran}

\bibliography{mybib}


\end{document}